\documentclass[aip,jcp,reprint]{revtex4-2}
\usepackage{mathptmx}
\usepackage[utf8]{inputenc}
\usepackage{amsmath}
\usepackage{graphicx}

\makeatletter

\providecommand{\tabularnewline}{\\}

\usepackage{braket}

\draft 

\makeatother

\begin{document}
\title{Velocity Gauge for Oscillator Strength in $\Delta$SCF theory}
\author{Yang Shen}
\thanks{These authors contributed equally to this work.}
\author{Yichen Fan}
\thanks{These authors contributed equally to this work.}
\affiliation{Department of Chemistry, Duke University, Durham, NC 27708}
\author{Weitao Yang}
\email{weitao.yang@duke.edu}

\affiliation{Department of Chemistry and Department of Physics, Duke University,
Durham, NC 27708}
\date{\today}
\begin{abstract}
Delta self-consistent-field ($\Delta$SCF) theory is widely used for
electronic excitation energy calculations. However, calculating the
corresponding oscillator strengths is challenging. The corresponding
many-electron wavefunctions are not directly accessible. Both the
ground-state and the excited-state wave functions from $\Delta$SCF
are described by reference Kohn-Sham (KS) single-determinant wavefunctions
for the fictitious non-interacting systems. The non-orthogonality
between the ground and excited Kohn-Sham determinants from two different
SCF calculations leads to unphysically origin-dependent transition
properties, such as transition dipole moment and length-gauge oscillator
strength. Including nuclei contribution in the perturbation is theoretically
rigorous, but its effectiveness is only limited to neutral systems,
as we show theoretically and numerically. While several other practical
approaches have been proposed to tackle the non-orthogonality problem
and yield reasonable results, inevitably the determinant of the ground
state or the excited state is changed, as well as the density matrix.
 In this work, we explore the use of the velocity gauge to compute
oscillator strength within $\Delta$SCF theory. We demonstrate that
the velocity gauge is capable of naturally accounting for the non-orthogonality
of $\Delta$SCF KS wavefunctions and offering origin-independent predictions
without any additional correction schemes to the KS wavefunctions.
Compared to the length-gauge results obtained via symmetric orthogonalization,
velocity gauge can offer comparable results. Furthermore, the adoption
of spin-purified singlet excitation energy in the velocity-gauge transition
dipole moment significantly enhances the overall performance of the
velocity gauge for $\Delta$SCF oscillator strength predictions on
conjugated chromophores.
\end{abstract}
\maketitle

\section{Introduction}

$\Delta$SCF within Kohn-Sham theory\citep{slater_statistical_1970,SLATER19721,ziegler_calculation_1977,RevModPhys.61.689,gilbert_self-consistent_2008,hait_highly_2020}
has proven effective in predicting excitation energies in both molecular
\citep{TRIGUERO1999195,cheng_rydberg_2008,besley_self-consistent-field_2009,kowalczyk_assessment_2011,doi:10.1021/jp5082802,doi:10.1021/acs.jctc.6b01161,amerikheirabadi_dft_2018,doi:10.1021/acs.jctc.7b00963,doi:10.1021/acs.jctc.7b00994,hait_excited_2020,doi:10.1021/acs.jpclett.1c02299,schmerwitz_variational_2022,10.1063/5.0075927}
andextended systems \citep{weissker_structure-_2004,zawadzki_absorption_2013,10.1063/5.0075927}.
Compared to methods like time-dependent density functional theory
(TDDFT)\citealp{runge_density-functional_1984,gross_local_1985,dreuw_single-reference_2005},
or equation of motion coupled cluster singles and doubles (EOM-CCSD)\citep{stanton_equation_1993},
$\Delta$SCF offers competitive accuracy with computational costs
comparable to ground-state DFT. In particular, $\Delta$SCF excels
in contexts where TDDFT struggles, such as core-level excitations
\citep{hait_highly_2020} and long-range charge transfer\citep{dreuw_failure_2004,dreuw_single-reference_2005},
often matching experimental results \citep{barca_simple_2018,hait_orbital_2021,lemke_highly_2024}.

Despite its practical success, the foundational principles of $\Delta$SCF
for excited state calculations had been ambiguous\citep{vandaele_scf_2022},
as the Hohenberg-Kohn theorem on the unique mapping of density to
the external potential or the many electron wavefunctions directly
does not extend to excited states. Recent theoretical advances have
established the theoretical foundation of $\Delta$SCF\citep{yang_foundation_2024}.
The newly established formulation is that the energy functional for
ground and excited states are the same functional, and has three sets
of equivalent basic variables describing the noninteracting reference
systems: (1) the excitation quantum number and the potential, (2)
the noninteracting Kohn-Sham wavefunction, or (3) the noninteracting
one electron reduced density matrix when the noninteracting Kohn-Sham
wavefunction is single-determinantal\citep{yang_foundation_2024}.
The minimum of the functional is the ground-state energy and, for
ground states, they are all equivalent to the Hohenberg-Kohn-Sham
method. The other stationary points of the functional provide the
excited-state energies and electron densities, establishing the foundation
for the \ensuremath{\Delta}SCF method as in practice. Further work
develops the linearity conditions for fractional charges in excited
state theory and the concepts of excited state chemical potentials
as the derivatives of the linear curves\citep{yang_fractional_2024}.
This in turn leads to the general chemical potential theorem: orbital
energies are chemical potentials in ground-state density functional
theory and excited-state \ensuremath{\Delta}SCF theory, therefore
approximating quasiparticle energies for ground states and also the
new concepts of excited state quasiparticle energies as charged excitations
from excited states\citep{yang_orbital_2024}.

While excitation energies have been well-studied, oscillator strength
predictions remain underexplored. Oscillator strength is critical
for optical spectroscopy, but hindered by the non-orthogonality between
$\Delta$SCF wavefunctions. Dating back to Slater's seminal work\citep{slater_statistical_1970,SLATER19721},
the non-orthogonality between $\Delta$SCF wavefunctions existed.
Some early explorations of solving Hartree-Fock equations for the
excited states apply Lagrange multipliers to ensure the virtual single-particle
orbital involved in the transition is orthogonal to all occupied state
of the ground state wavefunction\citep{hunt_excited_1969,huzinaga_virtual_1970,huzinaga_virtual_1971}.
It is also argued that enforcing the $\Delta$SCF wavefunctions to
be orthogonal to an approximate ground-state wave function is unnecessary
and undesired, propagating errors from the reference state to the
other wave functions\citep{gilbert_self-consistent_2008,besley_self-consistent-field_2009}.
However, this non-orthogonality leads to origin-dependence of transition
dipole moment, undermining the physical interpretability of transition
dipole moments\citep{gilbert_self-consistent_2008}. As shown in Eq.
\ref{eq:origin_dependence},

\begin{equation}
\begin{aligned} & \langle\Phi_{0}(\mathbf{R})|\hat{\mu}|\Phi_{m}(\mathbf{R})\rangle\equiv\sum_{i}^{N_{\text{elec}}}\int\mathrm{d}\mathbf{x}_{i}\langle\Phi_{0}(\mathbf{R})|\hat{x}_{i}|\Phi_{m}(\mathbf{R})\rangle\\
= & \sum_{i}^{N_{\text{elec}}}\int\mathrm{d}\mathbf{x}_{i}\langle\Phi_{0}(\mathbf{R})|(\hat{x}_{i}-\mathbf{R})|\Phi_{m}(\mathbf{R})\rangle+N_{\text{elec}}\mathbf{R}\langle\Phi_{0}(\mathbf{R})|\Phi_{m}(\mathbf{R})\rangle\\
= & \langle\Phi_{0}(\mathbf{0})|\hat{\mu}|\Phi_{m}(\mathbf{0})\rangle+N_{\text{elec}}\mathbf{R}\langle\Phi_{0}(\mathbf{0})|\Phi_{m}(\mathbf{0})\rangle,
\end{aligned}
\label{eq:origin_dependence}
\end{equation}
 a translation of the entire system by $\mathbf{R}$ can arbitrarily
change the transition dipole moment between the ground-state electronic
wavefunction $|\Phi_{0}\rangle$ and the excited-state electronic
wavefunction $|\Phi_{m}\rangle$, thus making it physically meaningless.
The vectors $\mathbf{0}$ and $\mathbf{R}$ in the parentheses indicate
that the position of the center of mass or the origin, at where the
electronic wavefunctions are solved. Other than the origin, the molecular
geometry remains the same.

Several strategies have been proposed to address this challenges.
Among these methods, non-orthogonal configuration interaction (NOCI)
\citep{thom_hartreefock_2009} and related work like NOCIS\citep{oosterbaan_non-orthogonal_2018},
improve descriptions of both the ground state wave functions and the
excited state wave functions, and corrects unphysical prediction on
the transition properties. However, the application to DFT is less
robust due to the delocalization error\citep{perdew_self-interaction_1981,mori-sanchez_localization_2008},
and additional assumptions required in methods like constrained DFT
configuration interaction (cDFT-CI)\citep{wu_configuration_2007}.

One class of methods enforces orthogonality via symmetric orthogonalization
on the ground and excited states \citep{bourne_worster_reliable_2021,lemke_highly_2024}.
This method preserves the translational symmetry of transition dipole
moment, and the orthogonality is retained by diagonalizing the $2\times2$
overlap matrix spanned by the ground-state and excited-state single
determinants. Nevertheless, inevitably this method creates different
ground states when performing orthogonalization for different pairs
of $\Delta$SCF ground and excited states.

Another way is to expand the determinant of the excited state from
$\Delta$SCF optimization into a linear combination of singly-excited
configurations constructed based on ground state calculation \citep{vandaele_scf_2022,vandaele_photodissociation_2022,malis_spinorbit_2022}.
In this way, the excited-state wave functions are orthogonalized against
the ground-state reference wave function. Although the exact wave
functions should be orthogonal to each other, the $\Delta$SCF wave
functions are the reference noninteracting system wavefunctions, which
are just used as approximations to the exact many-body wave functions.
Therefore, the $\Delta$SCF wavefunctions are not required to be orthogonal,
and such an orthogonalization may propagate the errors from the reference
state to the other states. \citep{gilbert_self-consistent_2008}

Another category of methods correct directly transition dipole moment
by adding terms to cancel the origin dependence. \citep{hait_highly_2020,hait_accurate_2020,hait_orbital_2021,carter-fenk_electron-affinity_2022,toffoli_accurate_2022,deshaye_electronic_2023}
As presented in the theory section, adding nuclei contribution (and
equivalent method, e.g. re-positioning the origin for neutral systems)
has solid theoretical foundation, and does solve the origin dependence
problem nicely for neutral systems. \citep{hait_accurate_2020,hait_highly_2020,hait_orbital_2021,toffoli_accurate_2022,deshaye_electronic_2023}
Nevertheless, for charged systems, the origin dependence is not canceled
completely. For the other methods in this category, the underlying
principle is rather vague\citep{carter-fenk_electron-affinity_2022}.

In this study, we develop a simple but effective solution: adopt the
velocity gauge to eliminate the origin dependence. Velocity gauge
is an exact but alternative way to the length gauge for the interaction
Hamiltonian between molecules and radiation fields. It has been used
in previous work for oscillator strength calculations, not directly
with the $\Delta$SCF KS wavefunctions, but with \emph{reconstructed}
excited state KS wavefunctions \citep{vandaele_photodissociation_2022,vandaele_scf_2022}\textbf{.
}The single determinant of the excited state is projected onto the
singly excited configurations generated from the ground state determinant.
The reconstructed excited-state determinant is then expressed as a
linear combination of those configurations, thereby no component of
the ground-state determinant itself remains\textbf{. }Nevertheless,
the $\Delta$SCF wavefunctions are reference wavefunctions for the
noninteracting systems and adiabatically connect to the corresponding
many-electron wavefunctions of the interacting systems for both ground
and excited states\citep{yang_foundation_2024}. They do not need
to be orthogonal by definition or from calculations. Without any sound
theoretical justification, it is best that we directly use them as
approximations to the wavefunctions of the interacting system. We
therefore retain the $\Delta$SCF wavefunctions as they are. This
approach is validated across small organic molecules and large conjugated
chromophores. Although it does not resolve non-orthogonality, it circumvents
its adverse effects in transition property calculations.

\section{Theory}

\subsection{Origin dependence in length gauge}

In the non-relativistic limit, the electronic Hamiltonian of a molecular
system interacting with an external electromagnetic field can be written
as Eq. \ref{eq:EM_hamiltonian}. When solving for the adiabatic electronic
states, the nuclei kinetic operator is separated:

\begin{equation}
\begin{aligned}\hat{H^{\text{el}}}= & \sum_{i}^{N_{\mathrm{elec}}}-\frac{1}{2}(\hat{p}_{i}+\mathbf{A}(\mathbf{r}_{i}))^{2}-\sum_{i}^{N_{\mathrm{elec}}}U(\mathbf{r}_{i})+\sum_{A}^{N_{\mathrm{nuclei}}}Z_{A}U(\mathbf{R}_{A})\\
 & +\sum_{A>B}^{N_{\text{nuclei}}}\frac{Z_{A}Z_{B}}{|\mathbf{R}_{A}-\mathbf{R}_{B}|}+\sum_{i>j}^{N_{\text{elec}}}\frac{1}{|\mathbf{r}_{i}-\mathbf{r}_{j}|}-\sum_{A}^{N_{\text{nuclei}}}\sum_{i}^{N_{\text{elec}}}\frac{Z_{A}}{|\mathbf{r}_{i}-\mathbf{R}_{A}|}.
\end{aligned}
\label{eq:EM_hamiltonian}
\end{equation}

The choice of the vector potential $\mathbf{A}(\mathbf{r})$ and the
scalar potential $U(\mathbf{r})$ is not unique, but rather gauge-dependent.
In the long wavelength limit, one may define the length gauge (LG)
as given in Eq.  \ref{eq:LG_EMfield}\citep{ding_gauge_2011},

\begin{equation}
\mathbf{A}_{\mathrm{LG}}(\mathbf{r},t)=\mathbf{0},\quad U_{\mathrm{LG}}(\mathbf{r},t)=-\mathbf{r}\cdot\mathbf{E}(t),\label{eq:LG_EMfield}
\end{equation}
in which $\ensuremath{\mathbf{E}(t)}$ represents the external time-dependent
electric field in the long-wavelength limit. The Hamiltonian perturbation
from the magnetic field  can be written as shown in Eq. \ref{eq:perturbation_length},
which is essentially the perturbation of electronic and nuclear electric
potentials,

\begin{equation}
\Delta\hat{H}_{\mathrm{LG}}(t)=\sum_{i}^{N_{\mathrm{elec}}}\mathbf{E}(t)\cdot\mathbf{r}_{i}-\sum_{A}^{N_{\mathrm{nuclei}}}Z_{A}\mathbf{E}(t)\cdot\mathbf{R}_{A}.\label{eq:perturbation_length}
\end{equation}

The Hamiltonian perturbation contributes to the off-diagonal coupling
between the ground state and the excited states.  As adopted in Franck
Condon factor derivation, both the ground state, $|\Psi_{0}\rangle$,
and the excited state, $|\Psi_{m}\rangle$, are full wavefunctions
(vibronic). These states may be expanded by the adiabatic states solved
from the electronic Hamiltonian, remaining exact.\citep{gu_discrete-variable_2023}
 Under Born-Oppenheimer (BO) approximation, the full wavefunction
can be written as

\begin{equation}
\Psi_{n}(\mathbf{r},\mathbf{R})=\Phi_{n}^{\mathrm{el}}(\mathbf{r};\mathbf{R})X_{n}(\mathbf{R}),
\end{equation}
in which $\Phi_{n}^{\text{el}}(\mathbf{r;\mathbf{R)}}$ represents
the electronic part of the n-th full wavefunction, and the coordinates
of the nuclei are parameters only for $\Phi_{n}^{\text{el}}(\mathbf{r;\mathbf{R)}}$.
$X_{n}(\mathbf{R})$ stands for the nuclear wavefunction. The transition
rate between these two states can be derived with Fermi's golden rule,
$w_{0m}=\frac{2\pi}{\hbar}|\langle\Phi_{0}|\Delta\hat{H|}\Phi_{m}\rangle|^{2}\delta(E_{0}-E_{m}+\omega)$.
For excited-state theory, such as time-dependent density functional
theory (TDDFT), $\Delta$SCF, and Bethe-Salpeter equation (BSE), typically
the excited-state calculation shares the same molecular geometry as
the ground state. This indicates the nuclei part of the many-body
state, $X_{n}(\left\{ \mathbf{R}\right\} )$, is approximated by a
Dirac delta function and independent of the index for adiabatic electronic
states, n. Therefore, the matrix element $\langle\Psi_{m}|\Delta\hat{H}_{\text{LG}}|\Psi_{0}\rangle$
is given in Eq. \ref{eq:LG_expansion},

\begin{equation}
\begin{aligned}\langle\Psi_{m}|\Delta\hat{H}_{\mathrm{LG}}|\Psi_{0}\rangle & =-\sum_{A}^{N_{\mathrm{nuc}}}Z_{A}\mathbf{E}\cdot\mathbf{R}_{A}\langle\Phi_{m}^{\mathrm{el}}|\Phi_{0}^{\mathrm{el}}\rangle+\langle\Phi_{m}^{\mathrm{el}}|\sum_{i}^{N_{\mathrm{elec}}}\mathbf{E}\cdot\mathbf{r}_{i}|\Phi_{0}^{\mathrm{el}}\rangle\end{aligned}
,\label{eq:LG_expansion}
\end{equation}

\noindent in accordance with Eq. 2 from Deshaye's work\citep{deshaye_electronic_2023}.
In TDDFT and BSE formalism, the first term in Eq. \ref{eq:LG_expansion}
is exactly zero due to the orthogonality condition, while the orthogonality
between ground state and excited states breaks in $\Delta$SCF. In
$\Delta$SCF, it is the absence of such a term leads to origin-dependent
behaviors of transition dipoles and oscillator strengths for neutral
systems. For neutral systems, the inclusion of the nuclei contribution
exactly cancels the origin dependence shown in Eq. \ref{eq:origin_dependence},
while in charged systems the origin-dependence is merely mitigated,
instead of completely canceled.

At the end of discussion on the length gauge, the definition of oscillator
strength is given in Eq.\ref{eq:LG_f} as a dimensionless physical
quantity, reflecting the probabilities of transition from ground state
to excited states, under the influence of electromagnetic field.

\begin{equation}
f_{\text{LG}}=\frac{2}{3}(E_{m}-E_{0})|\langle\Psi_{0}|\hat{\mu}|\Psi_{m}\rangle|^{2}.\label{eq:LG_f}
\end{equation}

\subsection{Velocity gauge}

The vector potential $\mathbf{A}_{\text{LG}}$ and scalar potential
$U_{\text{LG}}$ are not the unique solution to the Maxwell's equations.
Instead, the solution has a gauge dependence. Apart from the length
gauge, the scalar and vector potential in the velocity gauge are given
in Eq. \ref{eq:VG_EMfield}\citep{ding_gauge_2011},

\begin{equation}
\mathbf{A}_{\mathrm{VG}}(\mathbf{r},t)=-\int_{0}^{t}\mathbf{E}(t^{\prime})\mathrm{d}t^{\prime},\quad U_{\mathrm{VG}}(\mathbf{r},t)=0.\label{eq:VG_EMfield}
\end{equation}

In the velocity gauge, the perturbation to the molecular Hamiltonian
shown in Eq. \ref{eq:perturb_vg} in principle also includes the terms
describing nuclear kinetic operator coupled to the vector potenital,
which is omitted

\begin{equation}
\Delta\hat{H}_{\mathrm{VG}}(t)=-\frac{1}{2}\sum_{i}^{N_{\mathrm{elec}}}[\hat{p}_{i}\mathbf{A}(\mathbf{r}_{i})+\mathbf{A}(\mathbf{r}_{i})\hat{p}_{i}].\label{eq:perturb_vg}
\end{equation}

In a similar manner, the scattering cross section and oscillator strength
can be derived for the velocity gauge. The resulting equation for
the oscillator strength in velocity gauge from a detailed derivation\citep{cohen-tannoudji_quantum_1986,schatz_quantum_2002,lestrange_consequences_2015}
is provided below,

\begin{equation}
f_{\text{VG}}=\frac{2}{3}\frac{|\langle\Psi_{0}|\sum_{i}^{N_{\text{elec}}}\hat{p}_{i}|\Psi_{m}\rangle|^{2}}{E_{m}-E_{0}}=\frac{2}{3}\frac{|\langle\Phi_{0}|\sum_{i}^{N_{\text{elec}}}\hat{p_{i}}|\Phi_{m}\rangle|^{2}}{E_{m}-E_{0}}.
\end{equation}

Note that the commutator relation $[\hat{r},\hat{H}]=\mathrm{i}\hbar\hat{p}$
is not used in the derivation. Instead, the commutator relation can
be adopted to access the numerical equivalence of oscillator strength
in both gauges and the Thomas-Reiche-Kuhn sum rule (if the exact Hamiltonian
is adopted). The theoretical equivalence of oscillator strength in
the length gauge and velocity gauge is established on the gauge flexibility
of solutions to Maxwell's equations. Theoretically, the velocity gauge
inherits the translational invariance from the momentum operator $\hat{p}$,
which extends the validity of oscillator strength to charged systems
in $\Delta$SCF. Compared to the other length-gauge methods effective
for charged systems, switching to the velocity gauge avoids the origin
dependence problem without introducing any additional approximation
or assumption. Numerically, since the single-particle orbitals in
$\Delta$SCF ground state and excited states are solved from different
mean-field Hamiltonians, the numerical equivalence of length gauge
and velocity gauge in principle is not guaranteed and remains to be
explored in this work.

\subsection{Spin purification}

In $\Delta$SCF based on KS-DFT, the ground-state and excited-state
electron densities are generated by a single KS determinant. One well-known
deficiency of $\Delta$SCF theory is the spin contamination problem,
as a result of the single-determinant representation \citep{bagus_singlettriplet_1975,ziegler_calculation_1977}.
For molecules with a closed-shell ground state wavefunction, the open-shell
singlet excited states obtained with $\Delta$SCF optimization essentially
cannot be represented by a single Slater determinant. In the context
of single excitations, $\Delta$SCF for the excited states is usually
performed with the initial guess that one of the electrons is prompted
from an occupied (spin-)orbital, $i$, to an unoccupied (spin-)orbital,
$a$. The resulting wave function is described by a single determinant
($|\Phi_{0}^{i\alpha\rightarrow a\alpha}\rangle=\hat{c}_{a}^{\dagger}\hat{c}_{i}|\Phi_{0}\rangle$),
which is not a spin-pure state. The orbital relaxation in excited-state
$\Delta$SCF calculation may complicate the discussion, but since
typically the cross-spin relaxation is not allowed, the broken spin-symmetry
pertains. For simplicity, the orbital relaxation is ignored in the
discussion of spin symmetry, and the resulting configuration is a
mixture of singlet and triplet excited state, as shown in Eq. \ref{eq:mix_state}.

\begin{equation}
\begin{aligned}|\Phi_{0}^{i\alpha\rightarrow a\alpha}\rangle & =\frac{1}{\sqrt{2}}[\frac{1}{\sqrt{2}}(|\Phi_{0}^{i\alpha\rightarrow a\alpha}\rangle-|\Phi_{0}^{i\beta\rightarrow a\beta}\rangle]\\
 & ~+\frac{1}{\sqrt{2}}[\frac{1}{\sqrt{2}}(|\Phi_{0}^{i\alpha\rightarrow a\alpha}\rangle+|\Phi_{0}^{i\beta\rightarrow a\beta}\rangle]\\
 & =\frac{1}{\sqrt{2}}|T,M_{s}=0\rangle+\frac{1}{\sqrt{2}}|\mathrm{S\rangle}.
\end{aligned}
\label{eq:mix_state}
\end{equation}

In conventional $\Delta$SCF, obtaining $|\mathrm{T},M_{s}=0\rangle$
with the corresponding energy directly is difficult. The total energy
of the directly optimized state, as a functional of the non-interacting
one-electron reduced density matrix $\gamma_{s}$, is assumed to be
a mixture of the singlet energy and the triplet energy\citep{bagus_singlettriplet_1975,ziegler_calculation_1977,kowalczyk_assessment_2011},
as shown in Eq. \ref{eq:mix_energy}.

\begin{equation}
E[\gamma_{s}(|\Phi^{i\alpha\rightarrow a\alpha}\rangle)]=E_{\mathrm{mix}}=\frac{1}{2}E_{\mathrm{T},M_{s}=0}+\frac{1}{2}E_{\mathrm{S}}.\label{eq:mix_energy}
\end{equation}

Based on the degeneracy of $|\mathrm{T},M_{s}=0\rangle$ and $|\mathrm{T},M_{s}=\pm1\rangle$\citep{lischner_first-principles_2012,hait_orbital_2021},
$E_{\mathrm{T},M_{s}=0}$ can be accessed by $E_{\mathrm{T},M_{s}=\pm1}$,
which can be obtained by a $\Delta$SCF optimization of the triplet
state\citep{bagus_singlettriplet_1975,ziegler_calculation_1977,hait_orbital_2021}.
Therefore, Eq. \ref{eq:spin_purified} gives the spin-purified energy
of the singlet state \citep{bagus_singlettriplet_1975,ziegler_calculation_1977,yamaguchi_spin_1988,xu_testing_2014}.

\begin{equation}
E_{\mathrm{S}}=2E_{\mathrm{mixed}}-E_{\mathrm{T},M_{s}=0}=2E_{\mathrm{mixed}}-E_{\mathrm{T},M_{s}=\pm1}.\label{eq:spin_purified}
\end{equation}

The transition momentum in the velocity gauge might be computed by
Eq. \ref{eq:AP_wave}. For simplicity, $\hat{P}=\sum_{i}^{N_{\text{elec}}}\hat{p}_{i}$
is used in the following equations.

\begin{equation}
\begin{aligned}\langle\Phi_{0}|\hat{P}|\mathrm{S\rangle} & =\frac{1}{\sqrt{2}}\langle\Phi_{0}|\hat{P}|\Phi_{0}^{i\alpha\rightarrow a\alpha}\rangle+\frac{1}{\sqrt{2}}\langle\Phi_{0}|\hat{P}|\Phi_{0}^{i\beta\rightarrow a\beta}\rangle\\
 & =\sqrt{2}\langle\Phi_{0}|\hat{P}|\Phi_{0}^{i\alpha\rightarrow a\alpha}\rangle.
\end{aligned}
\label{eq:AP_wave}
\end{equation}

One may argue that it is the multi-configurational singlet state that
should be considered when computing the transition dipole moment in
the length gauge (or the transition momentum in the velocity gauge)\citep{vandaele_photodissociation_2022,vandaele_scf_2022}.
Nevertheless, such a multi-configurational singlet state is typically
not obtained by a direct $\Delta$SCF optimization, therefore the
spin-purified wavefunction may yield results that significantly deviate
from the many-electron wavefunction. An example of H$_{2}$ molecule
is provided and analyzed in the supporting information, which projects
the $\Delta$SCF excited-state wavefunction onto the singly excited
determinants and compares the contribution from each configuration
to the transition dipole moment to TDDFT. With unpurified $\Delta$SCF
single determinant, the transition dipole moment is in good agreement
with TDDFT. However, the procedure described in Eq. \ref{eq:AP_wave}
introduces an additional $\sqrt{2}$ pre-factor into the transition
dipole moment and deteriorates the results. In some specific scenarios,
the configuration interaction in DFT may be feasible by performing
cDFT-CI, but it introduces additional assumptions on the locality
of some occupied orbitals \citep{wu_direct_2005,wu_constrained_2006,wu_configuration_2007}.
To keep consistent with the literature on the comparison between $\Delta$SCF
oscillator strength and TDDFT (EOM-CCSD) oscillator strength \citep{bourne_worster_reliable_2021},
in this work the transition momentum (and transition dipole moment)
is computed directly with the $\Delta$SCF wavefunctions without purification.
While these single-determinant wavefunctions are approximations to
the exact wave functions\citep{hait_highly_2020,hait_accurate_2020},
they yield reasonable spectra\citep{hait_accurate_2020,hait_highly_2020,vandaele_photodissociation_2022}.

\subsection{Implementation}

The transition momentum (or transition dipole moment) between $|\Phi_{0}\rangle$
and $|\Phi_{m}\rangle$ is evaluated as shown in Eq. \ref{eq:transition_dipole},
in a wavefunction-like approach \citep{lowdin_quantum_1955}. Here,
$\psi_{i}^{0}$, $\psi_{j}^{m}$ are the indexes for the occupied
orbitals of the ground state and the excited state, and $\mathbf{S}^{0m}$
is the overlap matrix between the occupied orbitals of different states.
``adj'' denotes adjugate matrix.

\begin{equation}
\langle\Phi_{o}|\hat{P}|\Phi_{m}\rangle=\sum_{ij}\langle\psi_{i}^{0}|\hat{p}|\psi_{j}^{m}\rangle\mathrm{adj}(\mathbf{S}^{0m})_{ij}.\label{eq:transition_dipole}
\end{equation}

A unitary transform can be applied to the occupied orbitals of both
the ground state and the excited state respectively, to make the overlap
matrix $\mathbf{S}^{0m}$ diagonal, facilitating a straightforward
evaluation of the transition dipole moment at a computational cost
of $O(N^{3})$ \citep{thom_hartreefock_2009}. As a result of the
non-orthogonality between the ground state and the m-th excited state
obtained by $\Delta$SCF, the length-gauge transition dipole moment
loses its physical meaning. Since adding the nuclear contribution
is theoretically clear and remedies the origin dependence problem
of neutral systems only, it will always be considered for the length
gauge, unless specified particularly.

For exact theories, the equivalence of the length gauge and the velocity
gauge could be proved by the commutator relation $[\hat{r},\hat{H}]=\mathrm{i}\hbar\hat{p}$.
For $\Delta$SCF theory, the ground state and the excited states are
optimized separately, therefore the KS Hamiltonian itself is state-dependent,
which breaks the numerical agreement of different gauges. To demonstrate
this point and to facilitate a comparison to the reported data\citep{bourne_worster_reliable_2021},
the counterpart of transition dipole moment is defined as $\frac{\langle\Phi_{0}|\hat{P}|\Phi_{m}\rangle}{\Omega_{m}-\Omega_{0}}$.This
quantity will be referred as ``transition dipole moment'' directly
later. The adoption of neither the spin-purified excitation energy
nor the mixed singlet excitation energy would achieve identical results
to the length gauge (with nuclei correction). But for verification
purpose, the following two types of oscillator strength are calculated,
respectively in Eq. \ref{eq:velo_f_puri} and Eq. \ref{eq:velo_f_mix},
although some previous work supports the adoption of spin purified
excitation energy only\citep{vandaele_photodissociation_2022}.

\begin{equation}
f_{\text{VG}}^{\text{purified}}=\frac{2}{3}(\Omega_{m}^{\text{purified}}-\Omega_{0})|\frac{\langle\Phi_{0}|\hat{P}|\Phi_{m}\rangle}{\Omega_{m}^{\text{purified}}-\Omega_{0}}|^{2}.\label{eq:velo_f_puri}
\end{equation}

\begin{equation}
f_{\text{VG}}^{\text{unpurified}}=\frac{2}{3}(\Omega_{m}^{\text{purified}}-\Omega_{0})|\frac{\langle\Phi_{0}|\hat{P}|\Phi_{m}\rangle}{\Omega_{m}^{\text{unpurified}}-\Omega_{0}}|^{2}.\label{eq:velo_f_mix}
\end{equation}

\section{Computational details}

We tested small molecules and large conjugated chromophores previously
studied in the literature \citep{bourne_worster_reliable_2021}. The
lowest singlet excited states were computed using $\Delta$SCF with
the CAM-B3LYP functional\citep{yanai_new_2004} and aug-cc-pVTZ basis
set\citep{kendall_electron_1992,dunning_gaussian_1989,prascher_gaussian_2011}.
For those large conjugated chromophore molecules, the calculation
level was PBE0 functional\citep{perdew_generalized_1996,adamo_toward_1999}
with def2-SVP basis set\citep{weigend_balanced_2005}. The initial
guess for $\Delta$SCF calculation was constructed by prompting the
electron on the highest occupied molecular orbital (HOMO) to the lowest
unoccupied molecular orbital (LUMO), and MOM was adopted in the optimization
process. All calculations were performed using PySCF package\citep{sun_pyscf_2018,sun_recent_2020},
and the code for computing oscillator strength for $\Delta$SCF can
be found at https://github.com/yangshen24/deltascf\_f.

The results of $\Delta$SCF calculations were compared to the EOM-CCSD
and TDDFT results reported before\citep{bourne_worster_reliable_2021}.
For the majority of the tested molecules, the correspondence of excited
states obtained with diverse methods follows the original assignment\citep{bourne_worster_reliable_2021},
if without specification in the supporting information. The computed
$\Delta$SCF excitation energies are provided in the supporting information,
in excellent consistency with the previously reported data\citep{bourne_worster_reliable_2021}.
For the other molecules, the correspondence is established based on
the symmetry of excitations manually.

\section{Results}

\subsection{Performance on small molecules}

\begin{figure}
\includegraphics[width=1\linewidth]{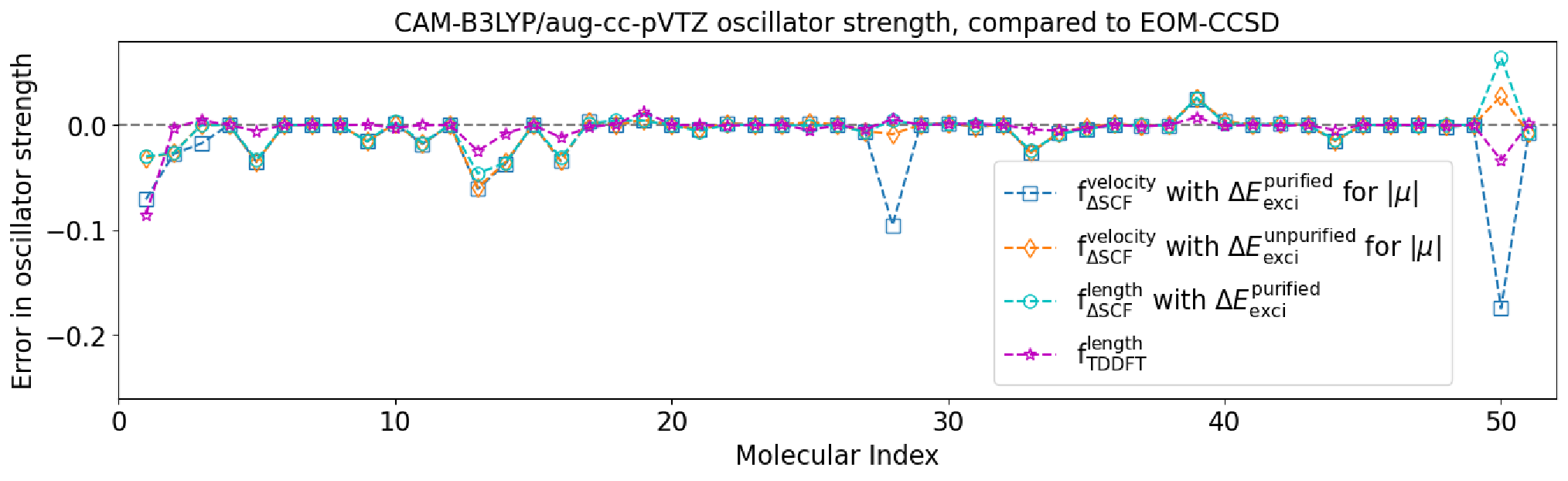}
 \caption{ The error in oscillator strength of the lowest singlet states. The
labels ``$\Delta\text{E}_{\text{exci}}^{\text{purified}}$'' and
``$\Delta\text{E}_{\text{exci}}^{\text{unpurified}}$'' indicate
the spin-purified singlet excitation energy or the unpurified singlet
excitation energy, which are used to calculate the corresponding transition
dipole moment (TDM), i.e. the denominator of $\frac{\langle\Phi_{0}|\hat{P}|\Phi_{m}\rangle}{\Delta E_{\text{exci}}}$.
The unpurified singlet excitation energy is defined as difference
between the ground-state and excited-state SCF ($\Delta$SCF) total
energy. The spin purification is performed according to Eq. \ref{eq:AP_wave}..
No additional correction on the wave function, e.g. symmetric orthogonalization
or projection, is adopted for the velocity gauge. All calculations
are done at the same DFT level (CAM-B3LYP/aug-cc-pVTZ), and the length-gauge
$\Delta$SCF results \citep{bourne_worster_reliable_2021} are based
on symmetric orthogonalization. The TDDFT results are in length gauge.
Connection lines are just guide to eyes.}\label{fig:small_f_error}
\end{figure}

We first tested the performance of the velocity gauge on the $\Delta$SCF
oscillator strength, for small molecules. Figure \ref{fig:small_f_error}
compares the oscillator strength in the velocity gauge ($f_{\text{unpurified}}^{\text{VG}}$,
$f_{\text{purified}}^{\text{VG}}$) to the results in length gauge,
with symmetric orthogonalization applied to the $\Delta$SCF ground
and excited states \citep{bourne_worster_reliable_2021}. Considering
nuclei contribution and performing symmetric orthogonalization in
length gauge shows very similar performance numerically, for neutral
molecules\citep{bourne_worster_reliable_2021}. The errors are computed
relative to the EOM-CCSD results, and a comparison to TDDFT results
is also provided\citep{bourne_worster_reliable_2021}. For the tested
51 molecules, $f_{\text{unpurified}}^{\text{VG}}$ give a prediction
comparable to the oscillator strength in length gauge with symmetric
orthogonalization. Since the splitting between the singlet state and
the triplet state is typically much less significant than the excitation
energy, in most cases $f_{\text{purified}}^{\text{VG}}$ performs
similarly to $f_{\text{unpurified}}^{\text{VG}}$. Only for tetracyanoethylene
and cis-2-butene (index 50 and 29 in Figure \ref{fig:small_f_error}),
the splitting of spin-purified singlet state and the triplet state
(more than 2 eV) significantly contributes to the oscillator strength
difference, therefore the results deviate more from the EOM-CCSD reference.
Overall, the performance of $f_{\text{purified}}^{\text{VG}}$ is
comparable to that of $f_{\text{unpurified}}^{\text{VG}}$.

Table \ref{tab:small_f} reports the mean absolute errors of both
transition dipole moment and the oscillator strength obtained with
these three methods. In the velocity gauge, the MAEs of both the ``transition
dipole moment'' and the oscillator strength obtained with mixed singlet
energy are almost identical to the $\Delta$SCF results in the length
gauge with symmetric orthogonalization. While the adoption of spin-purified
energy gives a slightly larger MAE, considering the corresponding
standard deviations, the difference is insignificant.

\begin{table}
\centering \caption{ The Mean Absolute Error (MAE) and the associated standard deviation
of oscillator strength ($f$) and the modulus of transition dipole
moment ($|\mu|$) computed with different methods, on a set of small
molecules . The rows of $\Delta|\mu|$ and $\Delta f$ represent the
standard deviations of the error associated with each method with
respect to EOM-CCSD results. The labels ``$\Delta E_{\text{exci}}^{\text{purified}}$''
and ``$\Delta E_{\text{exci}}^{\text{unpurified}}$'' indicate the
spin-purified singlet excitation energy or the unpurified singlet
excitation energy, which are used to calculate the corresponding transition
dipole moment (TDM), i.e. the denominator of $\frac{\langle\Phi_{0}|\hat{P}|\Phi_{m}\rangle}{\Delta E_{\text{exci}}}$.
The unpurified singlet excitation energy is defined as difference
between the ground-state and excited-state SCF ($\Delta$SCF) total
energy. The spin purification is performed according to Eq. \ref{eq:AP_wave}.
All calculations are done at the DFT level of CAM-B3LYP/aug-cc-pVTZ,
and the EOM-CCSD data is used as the reference\citep{bourne_worster_reliable_2021}.
All numbers are in atomic unit. Details on the oscillator strength
for each molecule can be found in the supporting information.}\label{tab:small_f}

\begin{ruledtabular}
\begin{tabular}{c|cc|cc}
 & \multicolumn{2}{c|}{$\Delta$SCF, velocity gauge} & \multicolumn{2}{c}{length gauge\citep{bourne_worster_reliable_2021}}\tabularnewline
\hline 
 & $\Delta\text{E}_{\text{exci}}^{\text{purified}}$ for $|\mu|$ & $\Delta\text{E}_{\text{exci}}^{\text{unpurified}}$ for $|\mu|$ & $\Delta$SCF, $\Delta\text{E}_{\text{exci}}^{\text{purified}}$ & TDDFT\tabularnewline
\hline 
$|\mu|$ & 0.06 & 0.05 & 0.05 & 0.02\tabularnewline
$\Delta|\mu|$ & 0.07 & 0.06 & 0.06 & 0.03\tabularnewline
$f$ & 0.014 & 0.008 & 0.008 & 0.0048\tabularnewline
$\Delta f$ & 0.030 & 0.013 & 0.014 & 0.013\tabularnewline
\end{tabular}
\end{ruledtabular}

\end{table}

\begin{figure}
\includegraphics[width=1\linewidth]{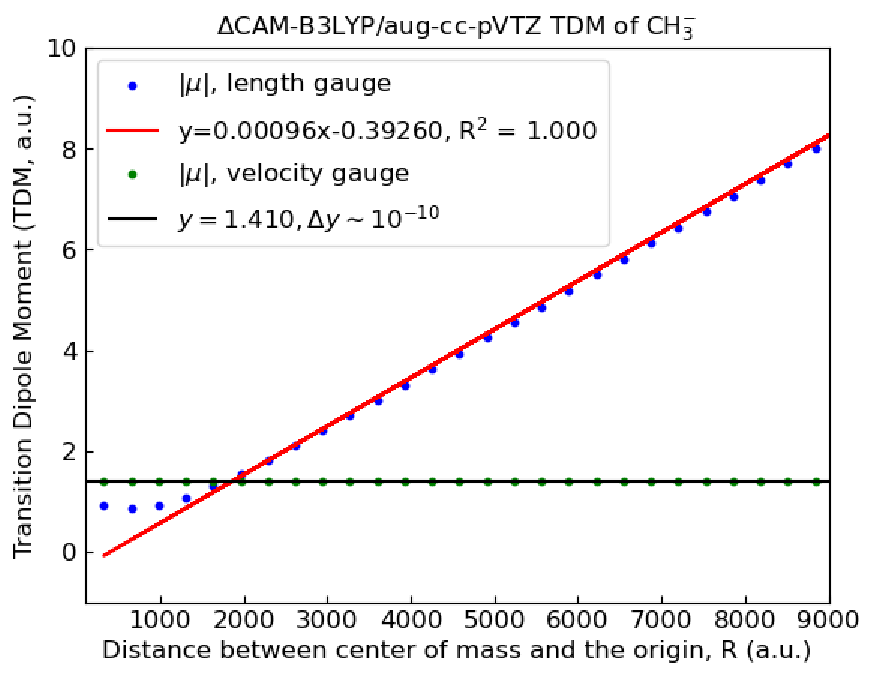}
\caption{ The origin dependence of length-gauge $\Delta$SCF transition dipole
moment with nuclei contribution and the origin-independence of velocity-gauge
$\Delta$SCF transition dipole moment, for a charged system, $\text{CH}_{3}^{-}$
anion. The blue and green points represents the modulus of transition
dipole moment in length gauge and velocity gauge, respectively. The
red line is a linear fit of the length-gauge transition dipole moment,
while the black line is for the velocity gauge. Since the transition
dipole moment is irrelevant to the origin in velocity gauge, only
the standard deviation of transition dipole moment in velocity gauge
is given, instead of the R$^{2}$.}\label{fig:CH3-}
\end{figure}

For neutral systems, adding the nuclei contribution to the perturbation
Hamiltonian yields origin-independent transition dipole moment in
the length gauge, as reported before \citep{bourne_worster_reliable_2021}.
For charged systems, the addition of nuclei correction cannot completely
cancel the origin dependence of transition dipole moment in length
gauge. For the tested small ions and anions, only the CH$_{3}^{-}$
anion shows a significant overlap between the ground state and the
first excited state, which is shown by the linear fit in figure \ref{fig:CH3-}.
The slope of length-gauge transition dipole moment w.r.t. seperation
is exactly the product of the net charge and the overlap between the
ground state and the excited state (0.00097). The standard deviation
of transition dipole moment in the velocity gauge is at the order
of $1\times10^{-10}$ a.u., which is negligible and can be attributed
to noise due to numerical precision. For $\Delta$SCF calculation
on charged systems, the velocity gauge is more efficient than the
length gauge. 

\subsection{Performance on conjugated chromophores}

\begin{figure}
\includegraphics[width=1\linewidth]{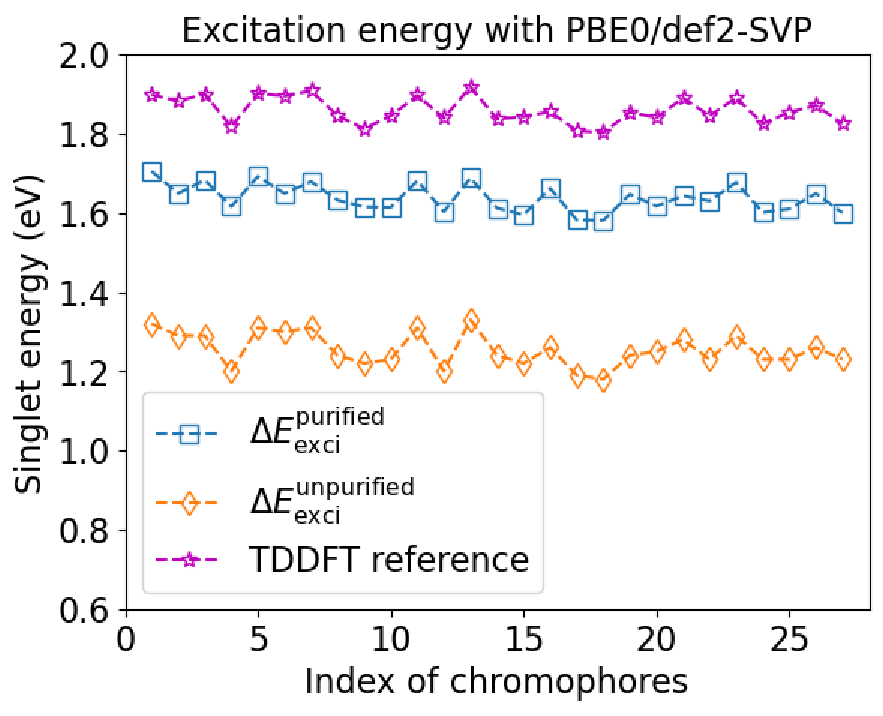} \caption{ The excitation energies of the chromophore lowest singlet states,
predicted by $\Delta\text{SCF}$ and TDDFT. The $\Delta$SCF calculation
is performed at the same DFT level (PBE0/def2-SVP) as the TDDFT reference
\citep{bourne_worster_reliable_2021}. Connection lines are just guide
to eyes.}\label{fig:chromo_ex}
\end{figure}

\begin{figure}
\includegraphics[width=1\columnwidth]{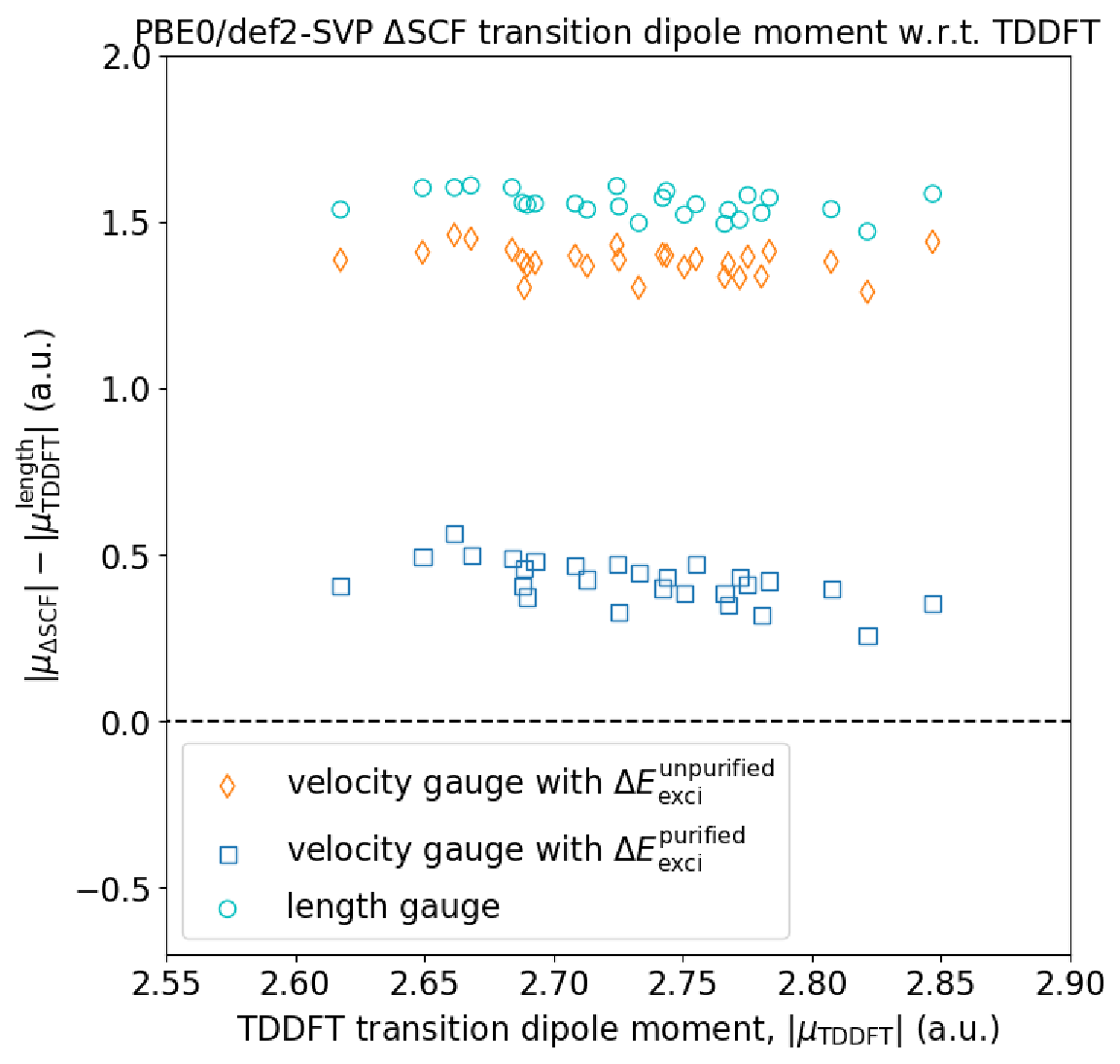}

\includegraphics[width=1\columnwidth]{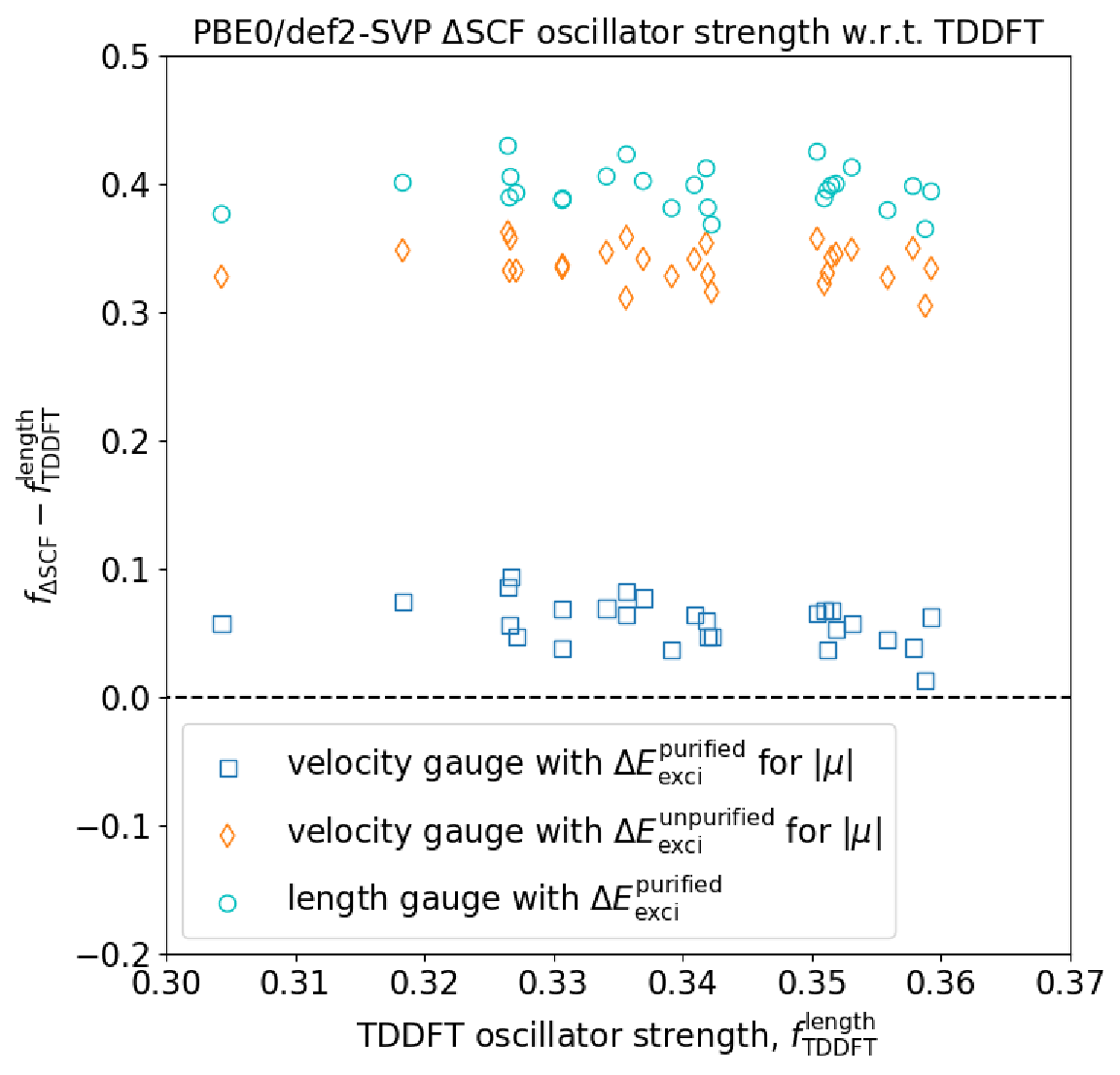}

\caption{ A comparison between the $\Delta$SCF transition properties computed
with different gauges and singlet excitation energies, at the DFT
level of PBE0/def2-SVP. The labels ``$\Delta E_{\text{exci}}^{\text{purified}}$''
and ``$\Delta E_{\text{exci}}^{\text{unpurified}}$'' indicate the
spin-purified singlet excitation energy or the unpurified singlet
excitation energy, which are used to calculate the corresponding transition
dipole moment (TDM), i.e. the denominator of $\frac{\langle\Phi_{0}|\hat{P}|\Phi_{m}\rangle}{\Delta E_{\text{exci}}}$.
The unpurified singlet excitation energy is defined as difference
between the ground-state and excited-state SCF ($\Delta$SCF) total
energy. The spin purification is performed according to Eq. \ref{eq:AP_wave}.
The results in length gauge adopting symmetric orthogonalization are
from \citep{bourne_worster_reliable_2021}. The black dashed line
($y=0.0$) is an indication of the agreement with the TDDFT oscillator
strength.}\label{fig:chromo_f}
\end{figure}

We move on to the tests of $\Delta$SCF transition properties on large
conjugated chromophores \citep{bourne_worster_reliable_2021}. For
these molecules, the splitting between the singlet state and the triplet
state is typically less than 1 eV, as shown in Figure \ref{fig:chromo_ex}.
Nevertheless, since the spin-purified singlet energy is less than
2.0 eV, the choice of spin-purified singlet energy or the mixed energy
significantly affects the results. An overview of the oscillator strength
computed for each chromophore by different methods is provided in
Figure \ref{fig:chromo_f}. For the tested chromophores, the results
computed with the velocity gauge are at least comparable to the reported
$\Delta$SCF data \citep{bourne_worster_reliable_2021}, if the mixed
energy is adopted. Although the excellent agreement to the previously
reported data is achieved, it should be clear that these results are
systematically and significantly overestimated, compared to the TDDFT
reference data.

The performance of spin-purified singlet energy in the calculation
of oscillator strength and transition dipole moment is also tested.
Due to the significantly splitting of singlet energy and triplet energy,
the ``transition dipole moment'' and oscillator strength obtained
are significantly reduced, therefore generating a much closer prediction
on these transition properties, which is evident in Figure \ref{fig:chromo_f}.

For all of the tested conjugated chromophore molecules, the prediction
of ``transition dipole moment'' and oscillator strength with spin-purified
singlet energy is significantly improved, as indicated by the mean
absolute percentage errors and the corresponding standard deviations
given in Table \ref{tab:large_f}.

\begin{table}
\caption{ The Mean Absolute Percentage Error (MAPE) and the associated standard
deviation of oscillator strength ($f$) and transition dipole moment
($|\mu|$) in diverse gauges. The rows of $\Delta|\mu|$ and $\Delta f$
indicate the standard deviation of the error associated with each
method with respect to the TDDFT reference, The reference data is
the TDDFT oscillator strength with the same DFT level calculation
(PBE0/def2-SVP)\citep{bourne_worster_reliable_2021}, and the $\Delta$SCF
calculation is performed at the same DFT level. The labels ``$\Delta E_{\text{exci}}^{\text{purified}}$''
and ``$\Delta E_{\text{exci}}^{\text{unpurified}}$'' indicate the
spin-purified singlet excitation energy or the unpurified singlet
excitation energy, which are used to calculate the corresponding transition
dipole moment (TDM), i.e. the denominator of $\frac{\langle\Phi_{0}|\hat{P}|\Phi_{m}\rangle}{\Delta E_{\text{exci}}}$.
The unpurified singlet excitation energy is defined as difference
between the ground-state and excited-state SCF ($\Delta$SCF) total
energy. The spin purification is performed according to Eq. \ref{eq:AP_wave}.
The unit is a.u. Details on the oscillator strength for each molecule
can be found in the supporting information.}\label{tab:large_f}

\begin{ruledtabular}
\begin{tabular}{c|cc|c}
 & \multicolumn{2}{c|}{$\Delta$SCF, velocity gauge} & length gauge\citep{bourne_worster_reliable_2021}\tabularnewline
\hline 
 & $\Delta\text{E}_{\text{exci}}^{\text{purified}}$ for $|\mu|$  & $\Delta\text{E}_{\text{exci}}^{\text{unpurified}}$ for $|\mu|$ & $\Delta$SCF, $\Delta\text{E}_{\text{exci}}^{\text{purified}}$\tabularnewline
\hline 
$|\mu|$ & 15.4\% & 50.6\% & 56.9\%\tabularnewline
$\Delta|\mu|$ & 2.5\% & 2.1\% & 2.1\%\tabularnewline
$f$ & 17.3\% & 99.5\% & 116.1\%\tabularnewline
$\Delta f$ & 5.5\% & 6.3\% & 8.0\%\tabularnewline
\end{tabular}
\end{ruledtabular}

\end{table}

\section{Discussion}

In this work, we developed the velocity gauge method in $\Delta$SCF
oscillator strength calculation and examined its performance for a
large set of organic molecules. Unlike TDDFT, in which the numerical
equivalence of length gauge and velocity gauge can be established
on the commutator relation between $\hat{r}$ and $\hat{H}$, $\Delta$SCF
does not possess an indentical KS Hamiltonian for the excited states
and the ground state). Therefore, in principle velocity gauge can
yield transition properties which differ significantly from length
gauge. Nevertheless, in all tested cases, the ``transition dipole
moment'' computed with the unpurified energy shows extremely high
agreement to the reported length-gauge data with symmetric orthogonalization.
This agreement strongly supports the effectiveness of adopting velocity
gauge in $\Delta$SCF oscillator strength calculation. Even for large
conjugated chromophore molecules, in which the $\Delta$SCF oscillator
strength in the length gauge is systematically overestimated, these
results are also replicated by the velocity gauge with unpurified
singlet excitation energy. With the help of spin-purified singlet
energy, the oscillator strength predicted in the velocity gauge gets
greatly improved, and the difference to the TDDFT reference data is
significantly reduced, at the expense of slightly worse prediction
on the set of small molecules. Nevertheless, we believe this error
should be acceptable, and overall adopting spin-purified singlet energy
achieves a better performance in the determination of $\Delta$SCF
oscillator strength.

It should be clear that this approach does not change the KS wave
functions, therefore the singlet excited state remains non-orthogonal
to the ground state. This non-orthogonal feature leads substantial
numerical difference between transition properties in the length gauge
and the velocity gauge. In length gauge, the non-orthogonality results
in origin-dependent and arbitrary transition properties. Although
the non-orthogonality can be remedied by a variety of methods, these
methods inevitably changes the determinants for the ground state or
the excited state (or both). However, from the perspective of adiabatic
connection\citep{yang_foundation_2024}, since the density matrix
of both states from $\Delta$SCF optimization can be mapped to the
corresponding many-body wavefunction, it is undesirable to change
the determinants. Our work shows that  this non-orthogonal feature
may not affect the physical meaning of oscillator strength in $\Delta$SCF
calculation, if the velocity gauge is adopted.

The usage of spin-purified singlet energy may indicate the spin-purified
wave function should also be used in the determination of ``transition
dipole moment'' and oscillator strength, as described in some other
work \citep{hait_accurate_2020,vandaele_photodissociation_2022,vandaele_scf_2022}.
Since the spin-purified wave function is not directly obtained from
$\Delta$SCF optimization, applying multi-configurational wave functions
for the singlet excited state may need more assumptions. Since overall
excellent agreement to the TDDFT is achieved by adopting $\Delta$SCF
in velocity gauge, for general $\Delta$SCF applications, we believe
that our method to be reliable  for predicting the oscillator strength,
which is intrinsically origin-independent. 

\section{Supplementary Materials}

See the supplementary material for all numerical data presented in
this paper. The geometries of the small molecules tested are also
provided.

\section{Acknowledgement}

We are grateful for the geometries of small molecules shared by Dr.
Susannah Bourne Worster from Durham University. Y.S., and W.Y. acknowledge
the support from National Institutes of Health (Grant No. 1R35GM158181-01)
\section{Author Declarations}

\subsection{Conflict of Interest}

The authors have no conflicts to disclose.

\subsection{Author Contribution}

Y.S. and Y.F. contributed equally to this paper.

\section{Data availability}

The data that support the findings of this study are available within
the article and its supplementary material.

\bibliography{updated_bibtex_file}

\end{document}